\documentclass[%
twocolumn,
 amsmath,amssymb,
 aps, physrev,
]{revtex4-2}

\usepackage{graphicx}
\usepackage{dcolumn}
\usepackage{bm}
\usepackage{textcomp}

\usepackage{siunitx}
\usepackage{hyperref}

\begin{document}

\title{Localized Thermometry via Dayem Bridges Integrated on Superconducting Qubit Chips}

\author{Ella O. Lachman}
\email{elachman@rigetti.com}
\affiliation{Rigetti Computing, 775 Heinz Avenue, Berkeley, CA 94710}

\author{Dave P. Pappas}
\affiliation{Rigetti Computing, 775 Heinz Avenue, Berkeley, CA 94710}

\author{Jayss Marshall}
\affiliation{Rigetti Computing, 775 Heinz Avenue, Berkeley, CA 94710}

\author{Josh Y. Mutus}
\affiliation{Rigetti Computing, 775 Heinz Avenue, Berkeley, CA 94710}

\date{\today}

\begin{abstract}
Accurate knowledge of the on-chip temperature is essential for understanding and optimizing the performance of superconducting qubits, yet direct thermometry at millikelvin temperatures remains challenging. While qubits themselves are sensitive to the temperature of their environment, other factors may affect the qubits' effective temperature, and using them as thermometers with any accuracy requires specialized measurement protocols and qubit designs, limiting their practicality for routine diagnostics and adding complex infrastructure to any hardware testing apparatus. Here we demonstrate a complementary on-chip thermometry method based on superconducting Dayem bridges that are integrated on the same chip as transmon qubits. By extracting the critical current of the Dayem bridge from I-V measurements, we obtain a local, quantitative measure of the chip temperature without the need for microwave calibration or qubit-specific control sequences. To demonstrate the utility of the Dayem bridges as thermometers, we fabricate them in-situ with qubits on the same chip, calibrate the Dayem bridge critical current as a function of temperature, and characterize its resolution and stability at cryogenic temperatures. We additionally perform simultaneous measurements of the Dayem bridge thermometer and qubit excited-state population, and show agreement over the relevant temperature range, validating the method against established qubit thermometry. Furthermore, we correlate the independently measured chip temperature with qubit energy relaxation and dephasing times, demonstrating the utility of this approach for diagnosing temperature-dependent decoherence mechanisms. These results establish integrated Dayem bridges as a simple, non-invasive, and scalable tool for cryogenic hardware development, and on-chip thermometry in superconducting quantum circuits.

\end{abstract}

\maketitle

\section{Introduction}
Accurate knowledge of temperature is essential for precise characterization of superconducting quantum circuits and proper error budgeting of high-fidelity operations. In transmon-based devices in particular, due to sensitivity to superconducting quasiparticle generation~\cite{Wilen2021CorrelatedCN, Martinis2009EnergyDI, Serniak2018HotNQ} and the lower frequencies at which they operate~\cite{Krantz2019QuantumEngineers, Koch2007Transmon}, temperature influences excited-state populations, quasiparticle densities, and coherence times, playing a central role in device performance and interpretation of experimental results. Even at millikelvin temperatures, small deviations from thermal equilibrium can measurably affect qubit properties \cite{jin_thermal_2015,yeh_microwave_2017}.
In practice, the temperature of a qubit chip often differs from the nominal temperature of the cryostat mixing chamber (MXC), usually measured by a $\text{RuO}_x$ resistive thermometer \cite{courts_commercial_2008}. The potential sources of this difference are many fold: imperfect thermalization, residual infrared radiation, and dissipation from control and readout wiring can lead to elevated local temperatures and spatial gradients across the chip. Consequently, standard cryostat thermometry does not necessarily reflect the temperature experienced locally by the quantum device.
Several approaches have been developed to probe temperature directly in superconducting circuits. 
While qubits themselves can serve as thermometers by extracting an effective temperature from the excited-state population or from multi-level protocols \cite{jin_thermal_2015,sultanov_protocol_2021}, this approach mandates the assumption that temperature is the only cause for excitations, and additionally introduces significant operational overhead. These methods rely on coherent control and calibration of the qubits and are ultimately limited by qubit coherence and quasiparticle-induced relaxation processes \cite{lvov_thermometry_2025}. Alternative approaches based on mesoscopic transport, including Coulomb blockade thermometry (CBT), provide primary temperature measurements through the conductance characteristics of tunnel junction arrays and have demonstrated high accuracy in the millikelvin regime \cite{pekola_thermometry_1994,shibahara_primary_2016,karkare_monte_2013}. However, such techniques typically require dedicated device structures and measurement schemes that are not naturally integrated into superconducting qubit platforms. These limitations motivate the development of simple, robust, and integrated thermometry techniques compatible with superconducting quantum circuits.

In this work, we investigate superconducting Dayem bridges (DBs) as on-chip thermometers. DBs are lithographically defined superconducting weak links (constrictions) that do not require an insulating tunnel barrier, distinguishing them from conventional Josephson junctions. Their properties have been studied since early investigations of thin-film superconducting bridges \cite{anderson_radio-frequency_1964, dayem_behavior_1967}, and more recently in contexts including field-effect superconducting devices, nanoscale interferometers, and weak-link sensors \cite{paolucci_ultra-efficient_2018,godfrey_investigation_2018,zgirski_nanosecond_2018}. Their simple geometry, compatibility with thin-film fabrication, and robustness to fabrication and handling make them attractive for integration into qubit devices.

Here we demonstrate that DBs can function as quantitative, on-chip thermometers when co-fabricated with superconducting qubits. While DB devices have previously been explored for thermometry and sensing \cite{zgirski_nanosecond_2018, zgirski_heat_2020, norowski_intrinsic_2026}, their application in superconducting quantum circuits introduces additional requirements, including sensitivity in the sub-\qty{100}{mK} regime relevant for qubit operation and compatibility with qubit fabrication and measurement workflows.

DBs offer several practical advantages that make them well suited for integration into superconducting qubit devices. First, they are straightforward to fabricate, requiring only a single electron-beam lithography step and a single superconducting thin-film deposition. Second, DBs are comparatively robust to subsequent fabrication processes. This allows them to be fabricated earlier in the process flow, while the sensitive qubit Josephson junctions can be patterned in the final fabrication step. In addition, the relatively large junction area of a DB and the fact that they are composed of a continuous metal trace, make the device less susceptible to electrostatic discharge during handling, packaging, and transport compared with conventional tunnel junctions. Lastly, at base temperature they are in the superconducting state, and thus do not risk being a parasitic channel for decoherence by introducing normal or lossy metals into the qubit device.

In addition to the ease of fabrication, the electrical measurement of the Dayem bridge can also be implemented with minimal experimental overhead compared to qubit measurements and other thermometry methods.
If qubit measurements are available in the system, DBs can be co-located directly on qubit chips because of the simplicity of the fabrication process and the compatibility of the materials with standard qubit device stacks. This enables simultaneous measurements of the bridge and the qubit within the same package and cryogenic environment, providing a direct probe of the local chip temperature experienced by the quantum device.

\begin{figure}
    \includegraphics[width=\columnwidth]{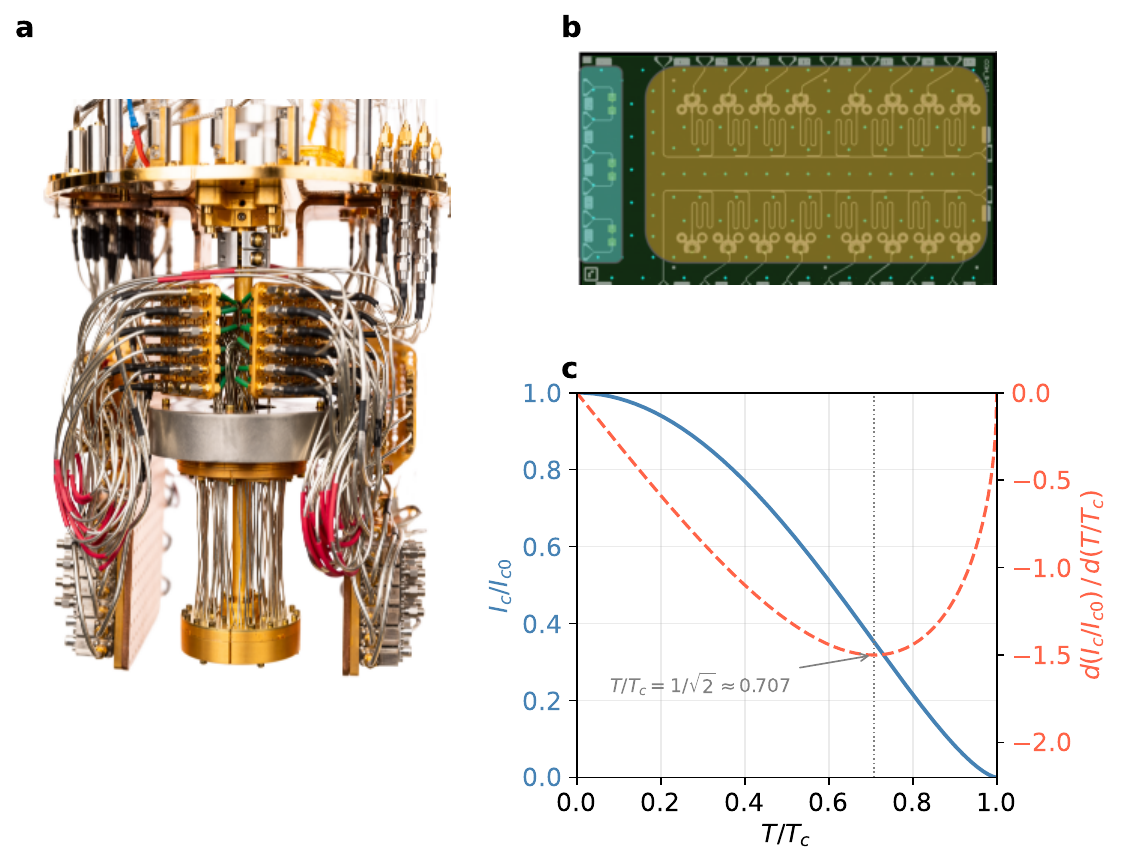}
    \caption{
    (a) An image of a standard Rigetti setup, illustrating the multiple interfaces and components between the MXC (fridge plate seen at the top of the image) and where the qubit chip is located (cylindrical package in the center, bottom of the image).
    (b) Chip schematic: qubits (highlighted yellow) and Dayem bridges (highlighted teal) on the same substrate in close proximity.
    (c) An illustration of the Dayem bridge's temperature-dependent critical current $I_c(T)$ (Eq.~\ref{eq:eq1}). In dashed lines, we also show the derivative of the dependence with an arrow marking the maximal point of the derivative. This point is the maximal temperature sensitivity of $I_c(T)$, where the temperature is equal to $1/\sqrt{2}$ of the critical temperature.
    }
    \label{fig:fig1}
\end{figure}

The operating principle of the DB thermometer relies on the temperature dependence of the superconducting critical current. For a short superconducting weak link, the critical current follows the Bardeen relation \cite{bardeen_critical_1962}

\begin{equation}
I_c(T) = I_c(0)\left[1-\left(\frac{T}{T_c}\right)^2\right]^{3/2},
\label{eq:eq1}
\end{equation}

where $I_c(0)$ is the critical current at zero temperature and $T_c$ is the superconducting transition temperature of the film. The critical current therefore decreases monotonically with increasing temperature, allowing the bridge to function as a thermometer. Knowledge of $T_c$ and $I_c(0)$ is sufficient to determine the device temperature from a measured $I_c$. Therefore, one can calibrate the thermometer or measure $T_c$ against a known reference thermometer and fit the resulting data to the Bardeen relation. An example of $I_c$'s dependence on temperature is shown in Fig.~\ref{fig:fig1}(c).

By measuring the critical current through current–voltage (I-V) measurements, we extract the local chip temperature without microwave calibration or qubit-specific protocols. We design and calibrate the thermometers to maximize sensitivity in the qubit operating regime, validate the extracted temperatures against direct qubit-based thermometry, and show that the independently measured temperature tracks variations in qubit coherence. These results establish integrated DB thermometers as a practical and scalable tool for temperature sensing in superconducting quantum circuits.

\section{Experimental Methods}
The devices studied in this work are planar superconducting circuits fabricated on high-resistivity silicon substrates. Each chip contains fixed-frequency and tunable transmon qubits capacitively coupled to coplanar waveguide resonators for dispersive readout. The qubit circuits are defined using standard lithographic techniques and thin-film deposition. Josephson junctions forming the nonlinear element of the transmons are fabricated using an aluminum shadow-evaporation process \cite{nersisyan2019manufacturinglowdissipationsuperconducting, Manenti2021}.

Devices are mounted to a printed circuit board inside a gold-plated copper package that is thermally anchored to the mixing chamber plate of a dilution refrigerator. The devices are shielded from magnetic fields with superconducting and cryoperm cans with a blackbody absorber to suppress stray infrared radiation. The input signal chain has a total attenuation of \qty{66}{dB} achieved by a series of attenuators anchored at different temperatures, and a \qty{7.65}{GHz} low-pass filter. The output signal line is filtered by isolators and amplified by a HEMT at \qty{4}{K} and a series of room-temperature amplifiers.
Single qubit gates were calibrated for 10 of the 16 transmons on the chip. For each qubit, $T_1$, $T_2$ Ramsey and readout fidelity measurements were performed at each temperature. Readout fidelity measurements were performed with $10,000$ shots in order to balance measurement time with the reliability of the effective temperature deduced from excited state population.

DB thermometers are integrated on the same chip as the qubits. Three bridges are fabricated on the chip in proximity to the qubits to probe the local temperature environment (see Fig.~\ref{fig:fig1}b). The bridges consist of narrow superconducting constrictions connecting two larger electrodes and are patterned using electron-beam lithography \cite{zgirski_nanosecond_2018}. Because the bridge geometry is defined lithographically and does not require a tunnel barrier, the fabrication process is relatively simple and robust. Following the metal deposition and patterning of the coplanar waveguides, resonators and qubit capacitive elements, the wafer undergoes the lithographically defined liftoff process to define the three thermometers.
As seen in Fig.~\ref{fig:fig1}c, the dependence of $I_c$ on temperature is roughly linear between $0.55\times T_c$ and $0.95\times T_c$, with the maximum at $1/\sqrt{2}\times T_c \approx 0.707\times T_c$. If we consider transmon operation temperature to be between \qty{10}{mK} and \qty{150}{mK}, that would put the desired $T_c$ for a thermometer material between \qty{160}{mK} and \qty{20}{mK}, where a lower $T_c$ also dictates a narrower sensitivity range. Previous tests with Titanium (Ti) thermometers yielded adequate results for a proof-of-concept, but the temperature sensitive area of the critical current was in the range of \qtyrange{150}{400}{mK} which is not low enough for transmon qubit applications. Palladium (Pd) was therefore selected to suppress the transition temperature ($T_c$) of the Ti layer via the inverse proximity effect, and push the temperature sensitive range to lower, more suitable temperatures. The material used for the DBs in this paper is therefore a Pd/Ti bilayer. These materials were selected to fit the fabrication process of the superconducting transmon devices, and to have the appropriate low critical temperature that would allow for sensitivity at the relevant temperature range. By varying the thickness of the Pd layer, we can suppress $T_c$ to the desired value. In this paper, we are presenting a single bilayer composition with a $1/10$ thickness ratio and a $T_c$ of \qty{166}{mK}.  

The DB thermometer critical current can be determined through $I$–$V$ measurements. A current bias is applied to the bridge and the voltage is monitored to detect the transition from the superconducting to the resistive state. The switching current extracted from these measurements is taken as the critical current $I_c$ of the device. In our setup, the bridge is embedded in a shunted measurement circuit and read out using a superconducting series array amplifier (SSAA)\cite{magnicon_squid_arrays}, which enables stable switching-current measurements while reducing the need for elaborate filtering typically required for four-wire transport measurements of superconducting devices. This method was previously used to track the critical current of a scanning SQUID device \cite{lachman_visualization_2015}. The current bias was applied via a twisted pair with a low-pass RC filter on the \qty{4}{K} stage.

At selected temperatures, a measurement of the $I_c$ is taken by ramping up the current ($I_{\text{in}}$) and monitoring the SSAA feedback voltage $V_{\text{SSAA}}$. SSAA feedback voltage can be converted to SSAA input current via a fixed factor derived from the SQUID input-to-feedback mutual inductance ratios and the feedback resistance setting. For each $I_{\text{in}}$ vs. $I_{\text{SSAA}}$ curve, the critical current $I_c$ is found by fitting linear regions in the superconducting and resistive branches and determining their intersection. An example of the data curves and the corresponding fitted critical currents is presented in Fig.~\ref{fig:fig2}(a).
After extracting the critical currents, we plot them vs. the MXC temperature. From the Bardeen equation (\ref{eq:eq1}), the critical current at $T=0$ and the critical temperature of the film $T_c$ are all that is needed in order to create the $I_c$ to $T$ conversion curve. We extract $T_c$ from independent measurements performed on a similar device in an Adiabatic Demagnetization Refrigeration (ADR) system with a current of \qty{10}{nA}. For $I_c(0)$ we use a fit to the Bardeen equation on data that was taken on a similar device when the device is directly mounted to the MXC, i.e not in the qubit package. In the future, the ADR system will also be used to independently extract $I_c(0)$ in the same way. With these values, we can get an independent estimate of the temperature that does not rely on the temperature of the MXC when the thermometers are combined with and cooled down in the same package as the qubits. The dashed line in Fig.~\ref{fig:fig2}b is not a fit to the data, but the Bardeen equation with the extracted values: $T_c=\qty{166}{mK}$ and $I_c(0)=\qty{23.34}{\micro\A}$.

Qubit characterization and thermometer measurements are performed within the same experimental cooldown cycle. The mixing chamber stage (MXC) temperature is set and stabilized using the cryostat's built-in PID control and heater. After reaching the temperature setpoint, there is a wait period of $>15$ minutes to allow the system to reach equilibrium. Next, current sweeps of the DB thermometers are acquired to determine the instantaneous critical current. Following the thermometry measurements, standard microwave pulse sequences are used to measure qubit energy relaxation and coherence times, as well as the excited-state population used for qubit-based thermometry. Because the thermometers are co-located with the qubits and measured in the same cryogenic environment, they provide a direct probe of the local temperature experienced by the quantum devices.

\section{Results}

\begin{figure}
    \includegraphics[width=\columnwidth]{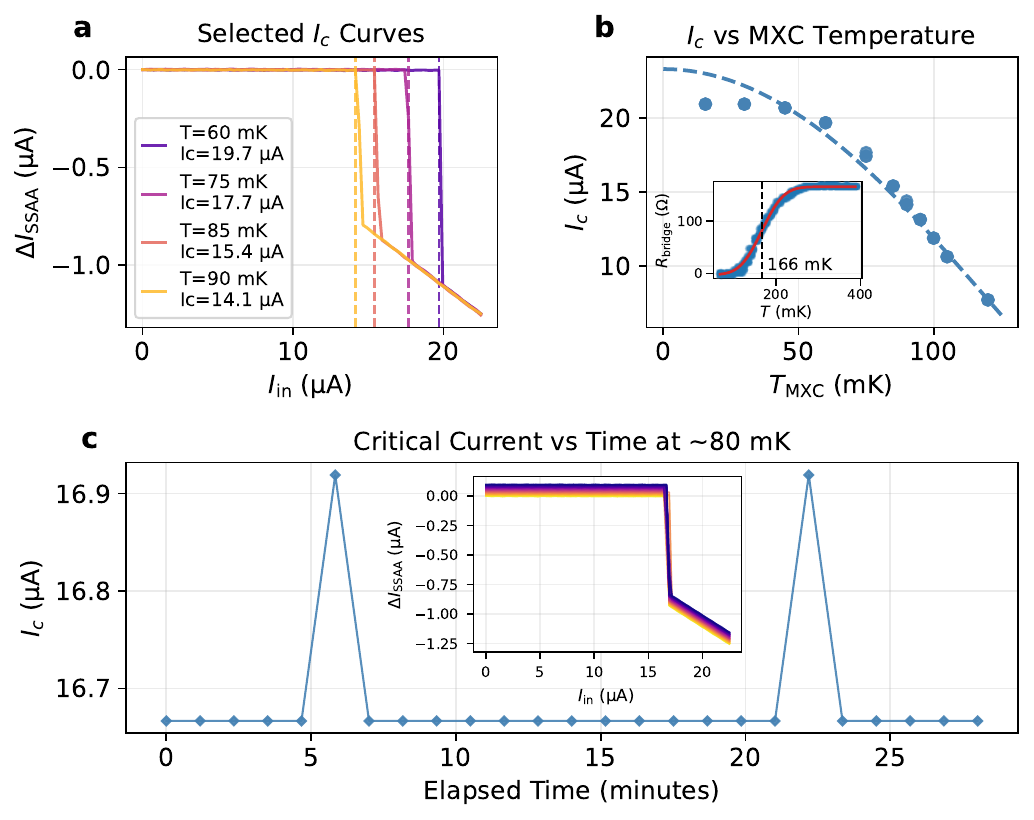}
    \caption{
    (a) Different Dayem bridge response curves for different temperatures. The low current flat branch is where no change is measured in the SSAA current, meaning all the current flows through the DB thermometer. The sharp jump is where $I_c$ is detected, followed by a resistive response where the majority of the current flows through the shunt resistor and the SSAA part of the circuit.
    (b) DB thermometer $I_c$ as a function of MXC temperature. Data (dots) and the Bardeen equation (dashed line) with the parameters: $T_c=\qty{166}{mK}$ and $I_c(0)=\qty{23.34}{\micro\A}$. Inset: independent $T_c$ measurement performed in an Adiabatic Demagnetization Refrigeration (ADR) system with a current of \qty{10}{nA}. The critical temperature extracted from the fit is $T_c=\qty{166}{mK}$.
    (c) $I_c$ stability over time. A half-hour-long measurement of the critical current of the DB thermometer at a high-sensitivity temperature point shows remarkable stability. The inset shows the response curves (similar to (a)), shifted for clarity, showing that the transition indeed happens at the same current for every measurement. The two outliers correspond to a temperature of \qty{72.93}{mK} compared to the baseline of \qty{74.44}{mK}, a difference of \qty{1.5}{mK}. We will note that all three thermometers were measured in an interleaved fashion and only one thermometer's data is presented here. The measurement rate for a single thermometer is therefore three times the one shown here.
    }
    \label{fig:fig2}
\end{figure}

As seen in Fig.~\ref{fig:fig2}(b), the data points below $\sim\qty{50}{mK}$ are showing a saturation and do not follow the predicted values for $I_c$. This indicates that the thermometers are not properly thermalized and therefore do not track the MXC temperature as they do above \qty{50}{mK}. There are two probable causes for this that we will test in future work: The first is heating as a result of the measurement itself or the twisted pair lines contacting the thermometers. The second is insufficient thermalization through the large pads defining the thermal bath for the electrons. In our initial design, these pads are only $(\qty{200}{\micro\m})^2$ whereas in most of the previous DB thermometry works these pads were much larger, and therefore allowed better thermalization of the electrons in the metal to the substrate before reaching the constriction (the DB itself). We will elaborate on this point in the discussion.

In order to show the stability of the measurement, we continuously measure $I_c$ over a period of 30~minutes. To ensure a stable thermal baseline, the MXC temperature is regulated under a PID feedback loop on the system heater throughout the acquisition window. Figure~\ref{fig:fig2}(c) shows that during that time, the measured curves show the same critical current with minimal deviation, confirming the stability and low variations in the measured quantity.

\begin{figure}
    \includegraphics[width=\columnwidth]{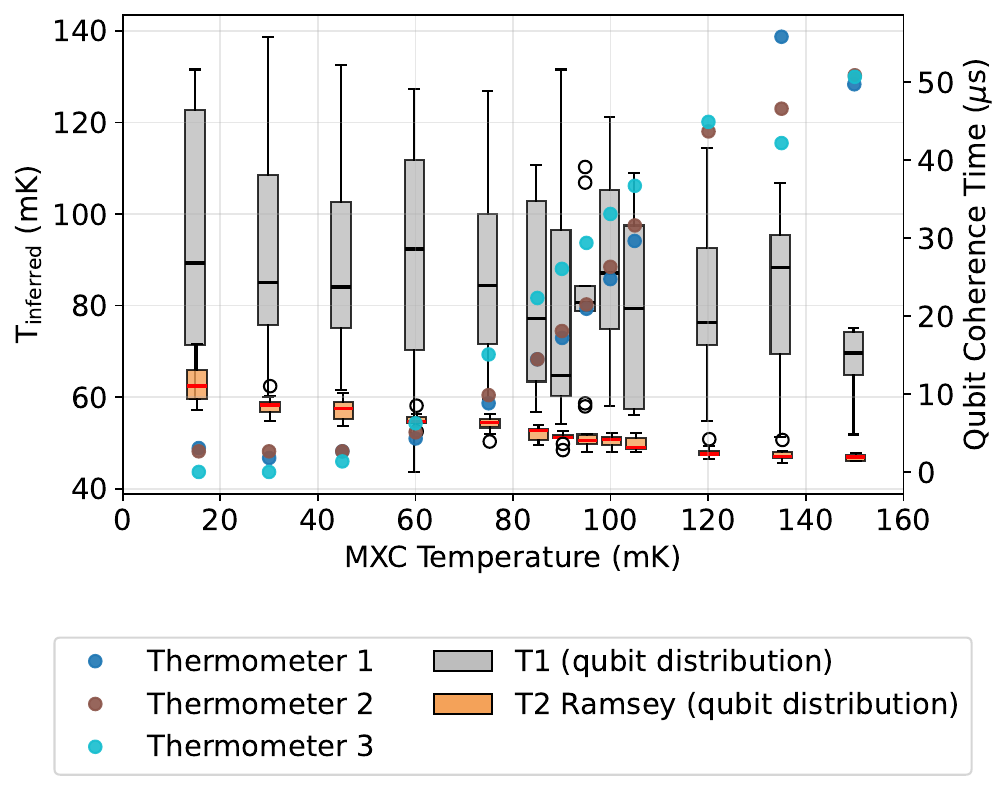}
    \caption{
    $T_1$ and $T_2$ Ramsey data from 10 qubits (right y-axis) and inferred temperature from 3 DB thermometers (left y-axis) as a function of fridge MXC temperature. Qubit data is presented as box plots, showing the median and the distribution of values. As expected, $T_1$ is less affected by increased temperatures than $T_2$ Ramsey.
    An inverse correlation is observed between the DB-inferred temperature and the $T_2$ Ramsey mean values; this trend confirms the utility of DB thermometers as a proxy for evaluating qubit coherence and dephasing.
    }
    \label{fig:fig3}
\end{figure}

Because these diagnostic devices are designed for rapid, efficient screening of qubit packaging during development, establishing a correlation between the temperature measured by the DBs and qubit relaxation and dephasing confirms that the DBs serve as a reliable proxy for overall qubit coherence performance. To demonstrate this, qubit coherence parameters were characterized immediately following each thermometry measurement at every stabilized temperature stage. From Fig.~\ref{fig:fig3}, it can be seen that both $T_1$ and $T_2$ Ramsey are affected by increased temperatures. The more affected qubit metric is $T_2$ Ramsey, with a decrease from a median of \qty{11.5}{\micro\s} at \qty{15}{mK} to a median of \qty{1.8}{\micro\s} at \qty{150}{mK}. In the range where the DB thermometers are sensitive to temperature changes (above \qty{50}{mK}), the temperature increase tracks the $T_2$ Ramsey decrease, showing that the DB thermometers can indeed be used as a proxy for temperature dependent qubit coherence when tested inside the same package.

\begin{figure}
    \includegraphics[width=\columnwidth]{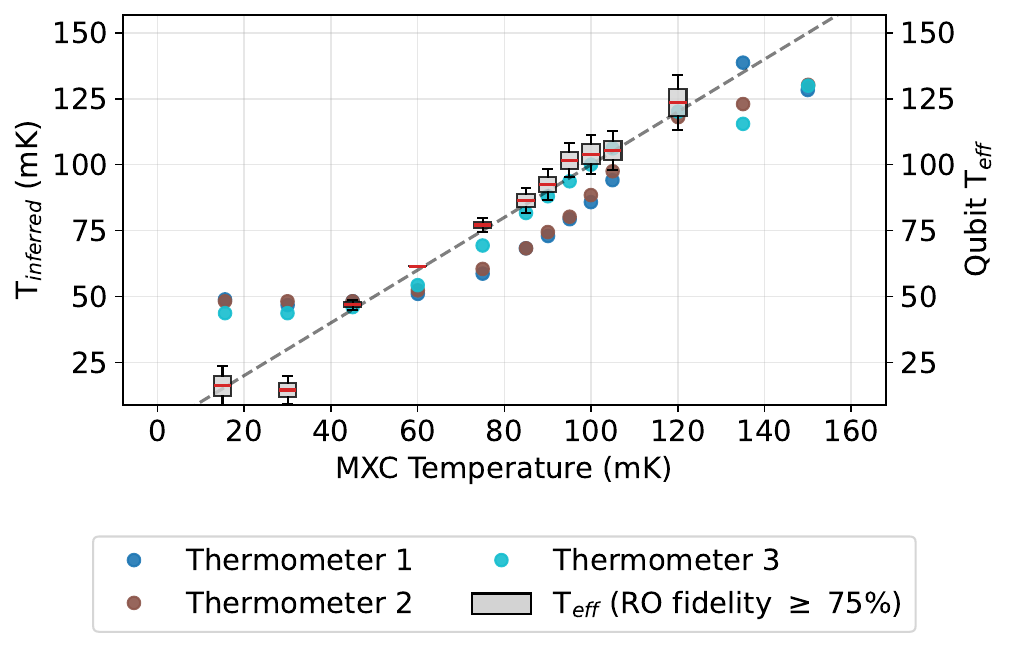}
    \caption{
    Effective qubit temperature from 2 qubits (right y-axis) and inferred temperature from 3 DB thermometers (left y-axis) as a function of fridge MXC temperature. Qubit data is presented as box plots, showing the median and the distribution of values.
    It can be seen that \textit{Thermometer 3} is following the MXC temperature as well as the qubits' effective temperature to lower temperatures than the other two thermometers. This can be attributed to a closer fit of the $I_c(0)$ used in the conversion curve (Fig.~\ref{fig:fig2}b) to the actual $I_c(0)$ of the device, and is plausible, as the room-temperature resistance of \textit{Thermometer 3} deviates from the other thermometer resistances.
    }
    \label{fig:fig4}
\end{figure}

In addition to $T_1$ and $T_2$ Ramsey, the qubits' effective temperature was also measured. This measurement uses the excited state population to estimate the effective temperature of the qubit through the Boltzmann distribution relation, $P_e/P_g = e^{-\hbar\omega/k_{\text{B}} T_{\text{eff}}}$, where $P_e/P_g$ is the ratio between the excited state population and the ground state population, and $\omega$ is the qubit $ \vert 0 \rangle \rightarrow \vert 1 \rangle$ transition frequency.
These measurements require a good readout fidelity and a number of shots large enough to properly sample the low-probability thermal transition to the excited state. Of the 10 qubits that were measured, only 2 qubits had readout fidelity $> 75\%$ at elevated temperatures, up to \qty{120}{mK}. Figure~\ref{fig:fig4} shows the effective temperature extracted from qubit excited state population and the temperature inferred from DB thermometers $I_c$ as a function of the MXC temperature. It can be seen that the selected qubits' excited state temperature matches well with the MXC temperature, showing that the standard package used by Rigetti for these chips does well to thermalize the chip. At temperatures where DB thermometer $I_c$ is responsive to temperature, it can be seen that the DB thermometer also agrees with both the qubits and the MXC thermometer, showing that the DB thermometers indeed capture the correct temperature. The variation between the thermometers would be studied in future work in order to minimize temperature error. The three thermometers are fabricated to be nominally identical, with room temperature resistances of \qty{608.86}{\Omega}, \qty{608.98}{\Omega} and \qty{609.57}{\Omega}.

\section{Discussion}
Extracting an effective temperature $T_{\text{eff}}$ from the excited-state population $P_e$ of a transmon qubit requires significant experimental overhead. As demonstrated in our results, accurate estimation of $T_{\text{eff}}$ via the Boltzmann distribution demands high readout fidelity ($>75\%$) and substantial statistical sampling ($10,000$ shots per point) in addition to the hardware complexity.

Moreover, when transmon qubits' performance deviates from predictions, their metrics alone cannot definitively isolate the root cause. While a degraded $T_2$ or an elevated excited-state population indicates a possible issue in the system, these signatures are ambiguous; they cannot distinguish an elevated local lattice temperature from other non-thermal error channels such as quasiparticles, flux noise, or stray infrared radiation. 
An independent, on-chip thermometer is therefore essential to decouple physical temperature from these competing decoherence mechanisms.

In addition, qubit measurements are often used for validating new packaging designs, resulting in large hardware overheads and longer testing times. Though qubits are definitely needed in the ultimate validation of packaging schemes, iteration can be accelerated using DB thermometers with simpler electronic setups for the initial stages. These simpler on-chip thermometers are beneficial as a way to streamline and speed up the validation process, and as a supplemental metrology tool for validation.

Dayem bridges, with the simple dependence of their critical current $I_c$ on temperature, have been demonstrated in this work to be suitable thermometers at superconducting qubits' relevant temperature range. The critical current has been shown to be stable, repeatable and constant under the measurement conditions, and our results demonstrate that these devices track both nominal MXC temperatures and qubit coherence trends. Another benefit is that the measurement itself requires considerably less cryogenic hardware than qubit measurements. All of these characteristics highlight the potential use of DBs as routine diagnostic tools for hardware development.

DBs are also easy to fabricate, requiring a single lithography step and a single superconducting layer, resulting in low-complexity metrology chips. These chips can conceptually host an array of thermometers, to map potential gradients across the chips, or across the package in a multi-chip architecture \cite{PhysRevApplied.21.054063}. 
In addition, as they do not introduce any dissipative elements onto the chip and can be incorporated early in the process flow, DBs can be fabricated directly onto qubit lattices and assist in discerning the root causes of qubit coherence degradation, and the study of temperature-dependent decoherence mechanisms.

There are two key limitations identified in this work. The first is the large (on the order of \qty{10}{mK}) deviation between the nominally identical thermometers. These are likely caused by variations in $I_c(0)$ or $T_c$ between the devices, which result in an imperfect calibration curve between $I_c$ and $T_{\text{inferred}}$. This can be resolved by higher fabrication repeatability, or a statistical study resulting in a prediction of $I_c(0)$ for each DB based on, e.g., room-temperature resistance. The second limitation identified is the flattening out of the critical current $I_c$ below approximately \qty{50}{mK} [Fig.~\ref{fig:fig2}(b)]. This saturation implies that the electrons within the weak link decouple from the MXC stage and pool at an elevated thermal floor. We attribute this to two primary mechanisms, which we intend to explore in a follow up study:
\begin{enumerate}
    \item \textit{Insufficient volume of the thermal bath:} In our current design, the cooling pads defining the thermal bath for the electron gas are limited to $(\qty{200}{\micro\m})^2$. Previous low-temperature DB architectures utilized significantly larger thermalization pads to enhance electron-phonon coupling to the underlying substrate, thermalizing the electrons before they reach the constriction itself. Expanding these geometries in future iterations should suppress this thermal bottleneck. An interesting study would be comparing different pad sizes, potentially as a proxy to assist in qubit design to maximize thermalization.
    \item \textit{Measurement-induced heating:} Dissipation introduced during the current sweep when the current surpasses $I_c$ and the DB is in its normal state. In order to avoid self-heating during the measurement, we may choose the more complicated pulsed measurement scheme shown by Zgirski et al.~\cite{zgirski_nanosecond_2018}, therefore reducing the duty cycle of the measurement. 
\end{enumerate}
    
\begin{figure}
    \includegraphics[width=0.8\columnwidth]{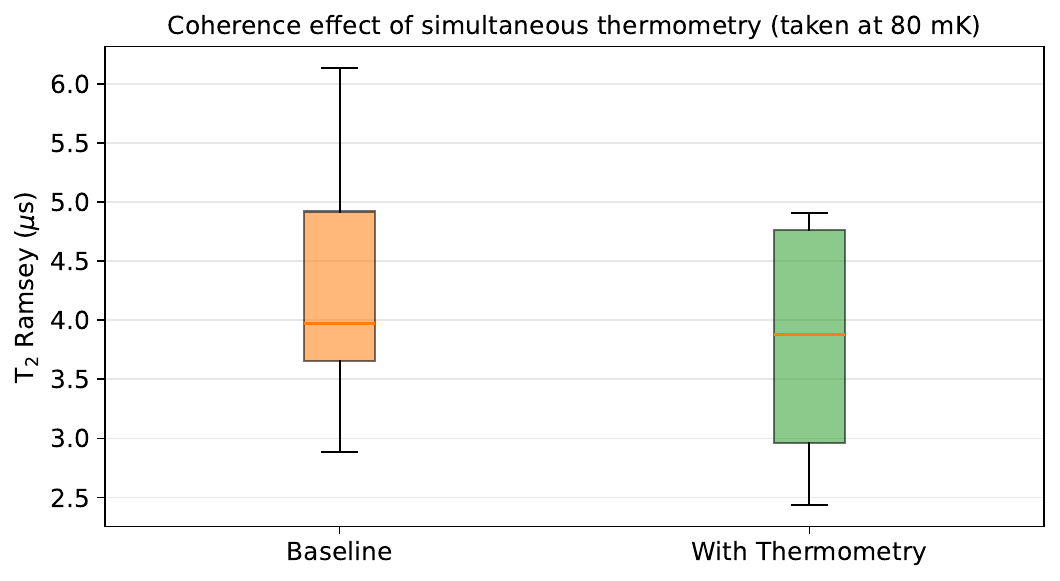}
    \caption{$T_2$ Ramsey distribution of a single qubit taken at \qty{80}{mK} acquired both actively during thermometer sweeps and independently as a baseline measurement. These are preliminary measurements with a modest sample size (baseline $n = 15$, thermometry $n = 8$). While a larger dataset is required for a definitive scaling analysis, these initial trials serve as an essential \textit{a priori} validation, demonstrating no immediate incompatibility or catastrophic crosstalk between the concurrent operations. Statistical comparison via Welch's $t$-test and Mann-Whitney $U$ test yields $p$-values well above $0.05$, indicating that the active thermometer readout does not introduce immediate, severe thermal or electronic penalties to adjacent qubit coherence.}
    \label{fig:fig5}
\end{figure}

Using the pulsed measurement method has the additional benefit of being faster, presenting opportunities for pump-probe experiments to diagnose the effects of localized heating on qubits. These could be induced by dissipation in the thermometers (if current passes $I_c$), or even diagnosing the thermal effects of control pulses on the qubit lattice. In Fig.~\ref{fig:fig5} we present limited statistical data on a single qubit taken with and without a concurrent measurement of the thermometers on the chip. Further study is merited, but our initial results show that the measurement of the thermometers does not significantly change the median $T_2$ Ramsey of the qubit under test.

With the combined ease of fabrication and relatively small footprint, one can imagine a thermometer next to every qubit on the chip, adding the much needed information of local temperature.

\section{Conclusion}
We have demonstrated that Dayem bridges provide a robust, quantitative, and low-overhead method for on-chip thermometry in superconducting quantum circuits. By engineering a proximity-coupled Pd/Ti bilayer film, we successfully depressed the sensor's critical temperature $T_c$ to shift its thermal sensitivity range directly into the sub-\qty{100}{mK} operating regime of transmon qubits. The extracted critical currents demonstrate excellent long-term stability and follow the predicted Bardeen relation down to approximately \qty{50}{mK}, tracking both nominal mixing chamber stage temperatures, qubit effective temperatures, and the physical trends of qubit decoherence.

Crucially, our preliminary cross-validation measurements show that these transport-based sensors can operate concurrently with quantum processors. Initial statistical analysis reveals no immediate degradation or severe crosstalk penalties in the $T_2$ Ramsey distribution of adjacent qubits during active thermometer sweeps. This operational compatibility, paired with a remarkably simple single-layer fabrication footprint and minimal cryogenic wiring infrastructure compared to traditional qubit thermometry, positions the Dayem bridge thermometer as a practical, scalable diagnostic tool for rapid prototyping and cryogenic packaging evaluation.

Future work will focus on expanding both the capabilities of these sensors and our understanding of local thermal dynamics. To overcome the observed electron-phonon decoupling bottleneck below \qty{50}{mK}, we intend to systematically investigate this phenomenon by varying the area of the pads. Additionally, shifting from continuous-wave current sweeps to nanosecond pulsed measurement architectures will drastically reduce the sensor duty cycle and mitigate self-heating. This faster readout paradigm opens up compelling opportunities for transient "pump-probe" style experiments. By utilizing the Dayem bridges to track microsecond-scale thermal recovery signatures following high-power microwave control or readout pulses, this integrated thermometry platform will offer deep insights into localized dissipation mechanisms, ultimately guiding the thermal design of next-generation, high-density superconducting quantum processors.

\begin{acknowledgments}

We would like to thank Stefano Poletto for critical reading of the manuscript, Joel Ullom for fruitful discussions, and the entire Rigetti team for providing general support and laying down the infrastructure that made this work possible.
E.L. conceptualized the project, coordinated efforts, performed the measurements and analysis in this work and wrote the manuscript.
D.P.P. performed $T_c$ measurements in the ADR system.
J.M. coordinated fabrication efforts.
J.Y.M. provided valuable insights and support.
All authors contributed to the manuscript.
This work was funded in part by NIST through the Quantum Economic Development Consortium (QED-C\textsuperscript{\scalebox{0.7}{\textregistered}}).
\end{acknowledgments}

\bibliographystyle{apsrev4-2}
\bibliography{references}

\end{document}